\documentclass[english,pra,twocolumn,superscriptaddress,showpacs,a4paper]{revtex4}
\usepackage[T1]{fontenc}
\usepackage[latin1]{inputenc}
\usepackage{graphicx}
\usepackage{amssymb}

\makeatletter

\makeatletter

\makeatletter

\makeatletter

\makeatletter

\usepackage{epsfig}

\usepackage{bm}

\newcommand{\identity}{\openone}

\makeatother

\makeatother

\makeatother

\makeatother

\usepackage{babel}
\makeatother

\begin{document}

\title{Analytic results on long-distance entanglement mediated by gapped
spin chains}

\author{Aires Ferreira and J. M. B. Lopes dos Santos }

\affiliation{CFP and Departamento de F\'{\i}sica, Faculdade Ciências Universidade
do Porto, 687 4169-007, Porto, Portugal}

\begin{abstract}
Recent numerical results showed that spin chains are able to produce
long-distance entanglement (LDE). We develop a formalism that allows
the computation of LDE for weakly interacting probes with gapped many-body
systems. At zero temperature, a dc response function determines the
ability of the physical system to generate genuine quantum correlations
between the probes. We show that the biquadratic Heisenberg spin-1
chain is able to produce LDE in the thermodynamic limit and that the
finite antiferromagnetic Heisenberg chain maximally entangles two
spin-$1/2$ probes very far apart. These results support the current
perspective of using quantum spin chains as entanglers or quantum
channels in quantum information devices. 
\end{abstract}
\maketitle
The quantum information (QI) approach of regarding entanglement as
a information \emph{resource} \citep{NMA00} stimulated important
developments in its characterization and in ways to measure it. On
the other hand, in condensed matter physics, quantum correlations
have long been recognized as an essential ingredient in many low-temperature
phases. These facts lead to a considerable amount of work on the characterization
of the entanglement properties of many-body systems at zero temperatures
(for reviews see \citep{ALFR}), particularly near quantum phase transitions
and also at finite temperatures \citep{ON,VMC,Ved}.

Feasible mechanisms of entanglement extraction from real solid state
\citep{DCG06} and their ability to transfer entanglement between
distant parties \citep{PMB05} are of crucial importance for the implementation
of QI protocols, such as teleportation or superdense coding. In systems
with short-range interactions, though, entanglement between two particles
usually decays quickly with distance between them \citep{ON}. However,
Campos Venuti \textit{et al} \citep{VBR06}, have recently found,
in numerical density matrix renormalization (DMRG) studies, that certain
correlated spin chains are able to establish long-distance entanglement
(LDE) between probes to which they couple, without the need of an
optimal measurement strategy onto the rest of the spins \citep{VMC}.
This naturally raises the question of which classes of strongly correlated
systems are able to produce LDE.

In this paper, we present a quantitative description of the effective
Hamiltonian of interaction between probes weakly coupled to gapped
many-body systems. This allows us to confirm some of the numerical
evidence of LDE found by DMRG \citep{VBR06} and to obtain new results
concerning two important spin chain models. Although our formalism
can be applied to general gapped many-body systems, here we focus
on one-dimensional spin chains. In the spirit of \citep{VBR06}, two
probes, $a$ and $b$, interact with the spin chain, locally, through
sites $m$ and $n$, respectively. The Hamiltonian of the system reads
$H=H_{0}+V_{m,n}$, where $H_{0}$ is the full many-body Hamiltonian
of the spin chain and $V_{m,n}=V_{a,m}+V_{b,n}$ describes the interaction
between the probes and the spin chain. We will show that, as long
as the probes interact weakly with the many-body system, the ground
state (g.s.) of the full system may display LDE between the probes,
\emph{i.e}., $E(\rho_{ab})>0$ when $|m-n|$ is of the order of the
system size. The opposite limit, \emph{i.e.,} strong interactions,
will cause the probes to develop robust correlations with the site
they interact inhibiting entanglement with them \citep{VBR06}. This
arises from a constraint on the correlations between different sub-systems
known as the monogamy of entanglement \citep{CKW00}. In order to
maximize the spin chain potential to entangle the probes, we require
that $J_{p}\ll J$, where $J_{p}$ is the interaction strength between
the probes ($a,b$) and the spins at $m$ and $n$, and $J$ is a
typical energy scale for the spin system (for instance, a nearest-neighbor
exchange interaction). For $J_{p}=0$ the state of the probes becomes
totally uncorrelated and the g.s. of the entire system becomes ($d\times d$)-fold-degenerate.
In this case we may write $|\psi\rangle=|\psi_{0}\rangle\otimes|\chi_{a}\rangle\otimes|\chi_{b}\rangle$,
where $|\psi_{0}\rangle$ is the g.s. of the spin chain (assumed non-degenerate)
and $|\chi_{\gamma}\rangle$ the state of the probe $\gamma$. The
role of the interaction ($J_{p}>0$) is to lift this degeneracy causing
the probes to develop correlations. For weak coupling $J_{p}$, the
effective Hamiltonian in this $d\times d$ low-energy subspace, obtained
%by integrating out the
as discussed below, will determine the dynamics and correlations of
the probes. For the special case of spin one-half probes ($d=2$),
the \textit{negativity} \citep{Vidal02}, or any other equivalent
entanglement monotone, can be used to quantify the entanglement.

\textit{The effective Hamiltonian.} In a very general way we can write
the local interaction between the probes and the corresponding sites
in the many-body system in the following manner: \begin{equation}
V=\sum_{\alpha=1}^{p}\gamma_{\alpha}^{a}O_{m}^{\alpha}\otimes A^{\alpha}\otimes\openone_{b}+\gamma_{\beta}^{b}O_{n}^{\alpha}\otimes\openone_{a}\otimes B^{\alpha},\label{eq:1}\end{equation}
 where $A^{\alpha}$$(B^{\alpha})$ denotes an operator acting on
the Hilbert space of the probe $a(b)$ and $\openone_{a(b)}$ the
corresponding identity operators. The many-body system operators on
sites $m$ are represented by $O_{m}^{\alpha}$ and $\gamma_{\alpha}^{a(b)}$
is the coupling strength. The projector onto the states with unperturbed
energy $E_{0}\equiv\langle\psi|H_{0}|\psi\rangle$ is $\mathcal{P}_{0}=|\psi_{0}\rangle\langle\psi_{0}|\otimes\openone_{a}\otimes\openone_{b}.$
Let $\mathcal{P}_{k}$ ($k>0)$ be the projector onto the subspace
of energy $E_{k}>E_{0}$ , so that $\openone=\mathcal{P}_{0}+\sum_{k>0}\mathcal{P}_{k}$.
Using the standard canonical transformation formalism \citep{SW66}
one can determine probes g.s. by diagonalizing an effective Hamiltonian
in the subspace spanned by $\mathcal{P}_{0}$, namely, $H_{eff}=\mathcal{P}_{0}(H_{0}+V)\mathcal{P}_{0}+|\psi_{0}\rangle\langle\psi_{0}|\otimes H^{(ab)}$
\citep{NOTE}. This a familiar concept that finds many applications
in condensed matter physics, such as, for instance, in the derivation
of the Ruderman-Kittel-Kasuya-Yosida magnetic interaction between
local moments in a metal \citep{KAS56b}. The coupling between the
probes is given by $H^{(ab)}=-\sum_{k>0}\langle UP_{k}U\rangle_{0}(E_{k}-E_{0})^{-1},$
where the average is taken with respect to the g.s. of the spin chain
$\langle UP_{k}U\rangle_{0}=\langle\psi_{0}|UP_{k}U|\psi_{0}\rangle$
and $U:=V-\langle V\rangle_{0}$. Note that, by definition, $\langle U\rangle_{0}=\mathcal{P}_{0}U\mathcal{P}_{0}=0$.
Entanglement between the probes arises from $H^{(ab)}$ since it contains
nonlocal terms such as $U_{a,m}\mathcal{P}_{k}U_{b,n}$ \citep{Li}.
The probe Hamiltonian can be transformed by straightforward manipulations
into an explicit form involving time dependent correlation functions
of the spin chain. A similar procedure is used to express cross sections
of scattering by many-body systems in terms of its correlation functions
\citep{Squ78}. We obtain (we set $\hbar=1$, see \citep{Correlation}
for the derivation) \[
H^{(ab)}=-\frac{1}{2\pi}\int_{-\infty}^{+\infty}\frac{dE}{E}\int_{-\infty}^{+\infty}dt\langle U(t)U\rangle_{0}e^{iEt}.\]
 We now introduce the explicit form of $U$ to arrive at the desired
result $H^{(ab)}=H_{L}^{(a)}+H_{L}^{(b)}+H_{NL}^{(ab)}+H_{L}^{(ab)},$
(see \citep{Local} for a comment on the local terms). Defining the
two-body connected correlation $\langle O_{m}^{\alpha}(t)O_{n}^{\beta}\rangle_{c}=\langle\psi_{0}|O_{m}^{\alpha}(t)O_{n}^{\beta}|\psi_{0}\rangle-\langle\psi_{0}|O_{m}^{\alpha}(t)|\psi_{0}\rangle\langle\psi_{0}|O_{n}^{\beta}|\psi_{0}\rangle$
the term coupling the two probes yields \begin{eqnarray}
H_{NL}^{(ab)} & = & \sum_{\alpha,\beta=1}^{p}\gamma_{\alpha}^{a}\gamma_{\beta}^{b}(C_{m\alpha;n\beta}+C_{n\beta;m\alpha})A^{\alpha}\otimes B^{\beta},\label{eq:2}\\
C_{m\alpha;n\beta} & = & \frac{1}{2i}\int_{-\infty}^{\infty}{dte^{-0^{+}|t|}sign(t)\langle O_{m}^{\alpha}(t)O_{n}^{\beta}(0)\rangle_{c}}.\label{eq:3}\end{eqnarray}
 The coupling between the probes can be expressed in terms of the
response function $\chi_{m\alpha;n\beta}(t)=-i\langle[O_{m}^{\alpha}(t),O_{n}^{\beta}]\rangle\theta(t)$,
where $\theta(t)$ is the Heaviside step function. Using the Lehman
representation at $T=0$ one can show that \begin{eqnarray*}
\tilde{\chi}{}_{m\alpha;n\beta}(0) & = & C_{m\alpha;n\beta}+C_{n\beta;m\alpha},\end{eqnarray*}
 where $\tilde{\chi}_{m\alpha;n\beta}(\omega)$ is the time Fourier
transform of $\chi_{m\alpha;n\beta}(t)$. The effective Hamiltonian
Eq.~(\ref{eq:2}) lifts the degeneracy of the GS level of the uncoupled
system ($J_{p}=0$). As long as the couplings appearing in $H_{NL}^{(ab)}$
are small compared to typical energy scales of the spin chain, such
as the gap to first excited state, $\Delta$, the low-energy physics
of this system, $E\ll\Delta$, with no real excitations of the spin
chain, will be well described by $H_{NL}^{(ab)}$. This condition
limits the strength of the chain-probe interaction, but is shown by
numerical results to be the appropriate limit to maximize LDE \citep{VBR06}.

In the remainder of the paper we will compute the LDE for two rotational
invariant spin chains: the finite Heisenberg spin-$1/2$ chain in
zero field using the exact results from bosonization theory; a specific
spin-1 Heisenberg chain with biquadratic interactions by means of
a approximation scheme for its spectrum. It is useful to write Eq.~(\ref{eq:1})
in terms of spin operators $\vec{S}_{m}$ for the spin chain and $\vec{\tau}_{a(b)}$
for the probes. Considering that the probes couple with the spin chain
via an Heisenberg interaction, the most common situation, $V_{m,n}=J_{a}\vec{S}_{m}\cdot\vec{\tau}_{a}+J_{b}\vec{S_{n}}\cdot\vec{\tau}_{b}$,
the connection with the previous notation becomes straightforward:
$O_{m}^{\alpha}=S_{m}^{\alpha}$, $A^{\alpha}(B^{\alpha})=\tau_{a(b)}^{\alpha}$
and $\gamma_{\alpha}^{a(b)}=J_{a(b)}$. Eq.~(\ref{eq:2}) becomes
simply, \begin{equation}
H_{eff}=J_{ab}\vec{\tau}_{a}\cdot\vec{\tau}_{b},\label{eq:H_eff_spins}\end{equation}
 where $J_{ab}=J_{a}J_{b}\tilde{\chi}{}_{m\alpha;n\alpha}(0)$ and
$\alpha=x$, $y$ or $z$.

\textit{The finite antiferromagnetic Heisenberg spin-$1/2$ chain.}
The isotropic antiferromagnetic Heisenberg model reads \begin{equation}
H:=\sum_{i=1}^{L-1}\vec{S}_{i}\cdot\vec{S}_{i+1}.\label{eq:Heisenbeg}\end{equation}
 Our formalism only applies to the \emph{finite} chain which has gapped
excitations. To calculate its time-dependent correlation functions
we will use the conformal invariance of the critical \emph{infinite}
chain ($L\rightarrow\infty$). We show below that its time-dependent
correlations are enough to extract the effective coupling $J_{ab}$
for the finite chain. It is clear that the effective Hamiltonian Eq.~(\ref{eq:2})
will preserve the full SU(2) symmetry of the interaction Hamiltonian
$H+V_{m,n}$, i.e. no local terms will give contribution to $H_{eff}$.
Assuming that the probes couple to the spin chain with the same strength
($J_{a}=J_{b}\equiv J_{p}$), $H_{eff}$ takes the very compact form
\begin{equation}
H_{eff}=J_{p}^{2}\tilde{\chi}_{m\alpha;n\alpha}(0)\vec{\tau}_{a}\cdot\vec{\tau}_{b}.\label{eq:eff_heisenberg}\end{equation}
 Hence, whenever $\tilde{\chi}_{m\alpha;n\alpha}(0)>0$ the g.s. of
the probes is a singlet-state displaying maximal LDE \citep{remark}.
Our computation of $\tilde{\chi}_{m\alpha;n\alpha}(0)$ will rely
on general bosonization results for correlations of spin-1/2 chains
(see \citep{TG04} for a review) and the conformal invariance of the
critical chain. The dominant long-distance correlations of the g.s.
of Hamiltonian Eq.~(\ref{eq:Heisenbeg}) oscillate with a $\pi$
phase change between neighbor spins. It is therefore useful to define
the retarded Green function for the staggered magnetization $M_{j}^{\alpha}:=(-1)^{j}S_{j}^{\alpha}$,
$G_{mn}^{R}(t):=i\langle[M_{m}^{\alpha},M_{n}^{\alpha}(t)]\rangle\theta(t)$.
The response function in Eq.~(\ref{eq:eff_heisenberg}) is $\chi{}_{m\alpha;n\alpha}(t)=(-1)^{|m-n|}G_{mn}^{R}(t)$.
The retarded Green function is obtained from the corresponding Matsubara
Green function, $G(x,\tau):=\langle\hat{T}_{\tau}M_{m}^{\alpha}(x_{m},\tau)M_{n}^{\alpha}(x_{n},0)\rangle$,
with imaginary time $\tau\in[-\beta,\beta]$ and where, in the limit
$|x_{m}-x_{n}|\gg1$, we may replace $x_{m}-x_{n}$ by a continuum
variable $x$. The Matsubara Green function for the infinite chain
reads \citep{SAC99} \[
G(x,\tau)=\mathcal{A}|v_{F}\tau+\imath x|^{-1};\]
 $\mathcal{A}$ is an amplitude and $v_{F}$ the Fermi velocity of
the spinon excitations. This result implies that the infinite chain
has a divergent $\tilde{\chi}_{m\alpha;n\alpha}(0)$; this is a direct
consequence of a zero gap and a signal of the critical nature of the
spin chain at $T=0$. Nevertheless, the finite chain is gapped, has
a finite $\tilde{\chi}_{m\alpha;n\alpha}(0)$, and the above result
can be used to calculate its correlation functions, precisely because
the critical nature of the infinite chain implies that $N$-point
correlation functions in different geometries are related by conformal
transformations \citep{Conformal}. The mapping of the infinite chain
to the finite chain is achieved by the following analytic transformation
(see Fig. 1), $w=\frac{L}{2\pi}\ln z=u+\imath r,$ where $z=v_{F}\tau+\imath x$.
\begin{figure}
\includegraphics[width=0.7\columnwidth]{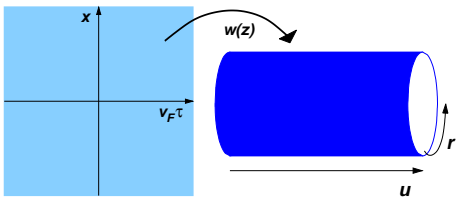}

\caption{The conformal transformation $w(z)=u+\imath r$ maps every point $(v_{F}\tau,x)$
in the plane into the strip geometry ($u\in]-\infty,\infty[$, $r\in[-L/2,L/2]$)
with periodic boundary conditions along the $v$ direction.}

\end{figure}

Using the transformation law for conformal invariant theories \citep{Conformal}
the Matsubara Green function for the finite antiferromagnetic Heisenberg
chain with periodic boundary conditions in the spatial coordinate
$r$ reads \citep{Cardy} $G_{cyl}(r,u)=2\pi\frac{\mathcal{A}}{L}(2\cosh(2\pi u/L)-2\cos(2\pi r/L))^{-\frac{1}{2}}.$The
analytic continuation to real time is made by Wick rotation $u\rightarrow\imath v_{F}t+0^{+}sgn(t)$
and the corresponding retarded Green function defined in the cylinder
$G_{cyl}^{R}(r,t)$ can be computed from the time-ordered Green function
$G^{R}(r,t)=-2\theta(t)\Im\left[G(r,t)\right]$(see \citep{TG04}).
Setting the branch cut of the logarithm in the negative real axis
we find

\[
G_{cyl}^{R}(r,t)=2\pi\frac{\mathcal{A}}{L}\frac{\theta(t)\theta(F(r,t))sign(\sin(2\pi t/L))}{\sqrt{F(r,t)}},\]
where $F(r,t)=2\cos(2\pi r/L)-2\cos(2\pi t/L)$. The response function
at zero frequency is then given by $\tilde{\chi}_{m\alpha;n\alpha}(0)=(-1)^{|m-n|}\int_{0}^{\infty}dtG^{R}(r,t)\exp(-0^{+}t$).
Here we only state the result,\[
\tilde{\chi}_{m\alpha;n\alpha}(0)=\mathcal{C}_{mn}\int_{2\pi r/L}^{\pi}dy\frac{y/\pi-1}{\sqrt{\cos(2\pi r/L)-\cos(y)}},\]
 with $\mathcal{C}_{mn}=(-1)^{|m-n|}\mathcal{A}/(2v_{F})$. Figure
2 shows the plot of the absolute value of the response function at
zero frequency confirming the existence of LDE for a wide range of
values of $r/L$. Note that $\tilde{\chi}_{m\alpha;n\alpha}(0)$ diverges
logarithmically at the origin. Our perturbative approach cannot be
applied unless $J_{p}^{2}\tilde{\chi}_{m\alpha;n\alpha}(0)\ll\Delta\sim J/L$,
and will fail in the thermodynamic limit ($L\to\infty$) for fixed
$r$. The numerical results of Ref. \citep{VBR06} show probes almost
completely entangled only for small values of $J_{p}\sim0.1$, for
a finite chain $L=26$. This value is well estimated by the limit
of validity of our perturbative approach, for $r/L\sim O(1)$, namely,
$J_{p}\ll J/\sqrt{L}$. These results, in the light of our analysis,
strongly suggest that the conditions for LDE are coincident with the
conditions for validity of the perturbative approach. Weakly coupled
probes get maximally entangled by the effective antiferromagnetic
interaction mediated by the spin chain. %
\begin{figure}
\includegraphics[width=0.7\linewidth]{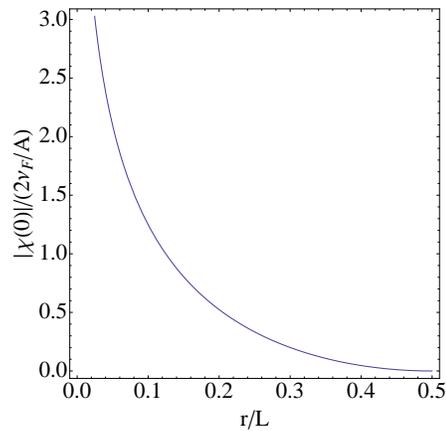}

\caption{The absolute value of the response function at zero frequency for
the finite antiferromagnetic Heisenberg chain. We have assumed $r\gg1$
so that the results from bosonization theory are accurate.}

\end{figure}

\textit{The AKLT model.} The Heisenberg spin one-chain with biquadratic
interactions reads\[
H:=\sum_{i=1}^{N-1}\left[\vec{S}_{i}\cdot\vec{S}_{i+1}+\beta(\vec{S}_{i}\cdot\vec{S}_{i+1})^{2}\right].\]
 This model admits an exact solution for $\beta=1/3$ which is known
as the Affleck-Kennedy-Lieb-Tasaki (AKLT) point \citep{AKL+88}. A
picture of the g.s. is given by the so-called valence-bond-solid (VBS).
Each spin-1 is represented by a couple of spins one-half, as long
as the antisymmetric state is projected out. The VBS state is constructed
by forming short-ranged singlets between nearest spin-$1/2$ and then
symmetrizing local pairs to get back $S=1$ states. In the thermodynamic
limit, the static correlations are very short-ranged {[}$\xi_{AKLT}=1/ln(3)\cong0.9$]
\citep{AKL+88}. For this reason we may ask whether two probes are
able to get entangled by interaction mediated by the spin-$1$ chain.
We cannot make an exact computation of LDE as in the Heisenberg model,
since the exact dynamical correlations are not known even for large
distances. However, as suggested by Arovas \textit{et al} \citep{ADP88},
we can apply the single-mode approximation (SMA) used to deduce the
phonon-roton curve in liquid 4He \citep{FRP72}, in order to study
the excitations in this model. This is done by assuming that a excited
state at wave vector $q$ is given by\[
|q\rangle\equiv S_{q}^{z}|\psi_{0}\rangle=N^{-1/2}\sum_{i}e^{\imath qr_{i}}S_{i}^{z}|\psi_{0}\rangle,\]
 where $|\psi_{0}\rangle$ is the exact g.s. of the AKLT model. Within
the SMA the dynamical structure factor $\mathcal{S}^{\alpha\beta}$
is related with the static structure factor defined as $s^{\alpha\beta}(q)=\langle\psi_{0}|S_{-q}^{\alpha}S_{q}^{\beta}|\psi_{0}\rangle$
in the simple way $\mathcal{S}(q,w)\cong s(q)\delta(w-w_{q})$. In
\citep{ADP88} it was shown that, $w_{q}=E_{q}-E_{0}=5(5+3\cos q)/27$
and that $s(q)=(10/27)(1-\cos q)/w_{q}$. The knowledge of the dynamical
structure factor allows us to compute the effective couplings of Eq.~(\ref{eq:2})
by inverse Fourier transform. For the AKLT model we obtain \[
\tilde{\chi}_{m+r\alpha;m\alpha}(0)=\int_{0}^{\infty}\frac{dt}{2\pi}\int_{-\pi}^{\pi}dq\cos(qr)\sin(w_{q}t)(a+\frac{b}{w_{q}})\]
 with $a=-2/3$, $b=80/81$. These integrals may be done by defining
$f(t)=\sin(w_{q}t)\theta(t)$ and noting that $\gamma=\int_{-\infty}^{\infty}dt\sin(w_{q}t)\theta(t)=\hat{f}(w=0)$
where $\hat{f}(w)$ is the Fourier transform of $f(t)$. We obtain
$2\hat{f}(w)=(w+w_{q}-i0^{+})^{-1}-(w-w_{q}-i0^{+})^{-1}$ and $\gamma=1/w_{q}$.
The remaining integral is done by extending the integrand to the complex
plain and computing the residues. This yields, $J_{ab}=-\frac{27}{10}J_{p}^{2}(-1)^{r}(1+\frac{4}{3}r)e^{-\frac{r}{\xi_{AKLT}}}$.
The sign of the interaction mediated by the AKLT spin chain changes
according to the distance between the probes. This comes from the
fact that the static correlations in this spin chain have a similar
alternation. Therefore at $T=0$ the probes get entangled whenever
their distance corresponds to a odd number of sites.

\textit{The effect of finite temperature and final comments.} If the
temperature is such that $k_{B}T\ll\Delta$, we do not expect real
excitations of the spin chain to be present. Only the subspace of
states described by $\hat{H}_{eff}$ will be populated and we may
calculate the correlations between the probes using $\rho_{ab}=e^{-\beta\hat{H}_{eff}}/Tr[e^{-\beta\hat{H}_{eff}}]$
with $\beta^{-1}=k_{B}T$. This defines a temperature, $1/\beta^{*}$,
above which entanglement disappears. For an antiferromagnetic $\hat{H}_{eff}$
the computation of the \emph{negativity} yields $\beta^{*}J_{mn}\simeq0.27$.
Loosely speaking, the probes will be entangled whenever the temperature
is smaller than the effective coupling between the probes. This formalism
has also been used to discuss qubit teleportation and state transfer
across spin chains \citep{VBR06}. In conclusion, we have expressed
the capacity of a gapped many-body system as an entangler of weakly
coupled probes at arbitrary distances in terms of a zero temperature
response function. We exemplified this formalism by calculating this
function for two quantum spin chains, shedding light on recent numerical
results on LDE. In this context, our results strongly suggest that
the main mechanism of LDE mediated by gapped spin chains is the existence
of dominant antiferromagnetic correlations. \textit{Acknowledgments.}
We gratefully acknowledge very enlightening discussions with J. Penedones.
A.F. was supported by FCT (Portugal) through PRAXIS Grant No. SFRH/BD/18292/04.

\bibliographystyle{apsrev}

\begin{thebibliography}{10}
\bibitem{NOTE}$\mathcal{P}_{0}H_{0}\mathcal{P}_{0}$ contributes
with a constant, and thus will be set to zero as it does not change
the eigenstates.
\bibitem{Correlation}We rewrite $H_{ab}$ as $H^{(ab)}=-\int_{-\infty}^{+\infty}\frac{dE}{E}\sum_{k>0}\langle\psi_{0}|UP_{k}U|\psi_{0}\rangle\delta(E-E_{k}+E_{0})$
and use the integral representation of the Dirac delta function, $\delta(E)=(2\pi)^{-1}\int_{-\infty}^{+\infty}dte^{iEt},$
and the fact that $\langle\psi_{0}|e^{iE_{0}t}Ue^{iE_{k}t}P_{k}U|\psi_{0}\rangle=\langle\psi_{0}|U(t)P_{k}U|\psi_{0}\rangle,$
where $U(t)=e^{iH_{0}t}Ue^{-iH_{0}t}$ is the probe-spin-chain coupling
in the Heisenberg representation for the spin chain. Finally we can
set $\langle U\rangle_{0}=0$ to our advantage, by including the term
$k=0$ in the sum over $k$ and, since $\openone=\sum_{k}\mathcal{P}_{k}$
, we obtain the stated result.
\bibitem{Local} The local terms in $H^{(ab)}$ read $H_{L}^{(ab)}=\sum_{\alpha,\beta=1}^{p}(\gamma_{\alpha}^{a}\gamma_{\beta}^{a}C_{m\alpha,m\beta}A_{\alpha}A_{\beta}\otimes\identity_{b}+\gamma_{\alpha}^{b}\gamma_{\beta}^{b}C_{n\alpha,n\beta}\identity_{a}\otimes B_{\alpha}B_{\beta})$
and $H_{L}^{a(b)}=\sum_{\alpha=1}^{p}\gamma_{\alpha}^{a(b)}A(B)^{\alpha}\langle O_{m}^{\alpha}\rangle$.
These terms are either constants or zero in the cases we consider.
\bibitem{Conformal} In a conformally invariant two-dimensional system
with scaling dimension $\eta$ the 2-point correlation function has
the transformation law $G_{g'}(w_{1},w_{2})=|dw(z_{1})/dz_{1}dw(z_{2})/dz_{2}|^{-\frac{\eta}{2}}G_{g}(z_{1},z_{2})$
generated by any analytic function $w(z)$. The subscripts $g$ and
$g'$ refer to the boundary geometry where the theory is defined.
\end{thebibliography}

\end{document}